# MAN IN THE MIDDLE AND CURRENT INJECTION ATTACKS AGAINST THE KLJN KEY EXCHANGER COMPROMISED BY DC SOURCES

MUTAZ Y. MELHEM [§] AND LASZLO B. KISH

*Department of Electrical and Computer Engineering,*
*Texas A&M University*
*College Station, TX 77843- 3128, USA*

*yar111@tamu.edu*

This study addresses a new question regarding the security of the Kirchhoff-Law-Johnson-Noise (KLJN) scheme compromised by DC sources at Alice and Bob: What is the impact of these parasitic sources on active attacks, such as the man-in-the-middle (MITM) attack, or the current injection attack? The surprising answer is that the parasitic DC sources actually increase the security of the system because, in the case of the MITM attack, they make easier to uncover the eavesdropping. In some of the cases Eve can fix this deficiency but then the problem gets reduced to the original MITM attack to which the KLJN scheme is immune, it is already as proven earlier.

*Keywords:* unconditional security; secure key exchange; KLJN; active attacks; DC ground loop.

## 1. Introduction

Information theoretic (IT, or unconditional) security [1,2], in the ideal case, does not allow the eavesdropper (Eve) to extract any information from the ciphertext about the plaintext message. Even though practical situations are never ideal thus, there is always some information leak, the leak can be made negligible by privacy amplification software tools [3] provided the bit error probability is sufficiently small.

IT security is future proof because, even if Eve records the message and the key exchange data, no future technical developments (such as quantum computers or noise-based logic processors) can help to extract information from the recorded data. This feature is a crucial advantage over conditional security that offers zero privacy when there is enough time for the computation or sufficiently large computational power becomes available in the future. Conditional security is not future proof.

One of the essential elements of IT security is the key exchange that must also be IT secure. This is a difficult goal because secure key exchange is itself a secure communication.

One example for achieving this goal via physical means and for the difficulties involved with this task is quantum key distribution (QKD) [3-40].

Another example are the Kirchhoff-Law-Johnson-Noise (KLJN) secure key distribution schemes and related systems [41-58] representing the sole class of *classical physical* protocols that provides IT security via statistical physics. This is the only system

---

[§] Corresponding author.

*Active Attack against the KLJN scheme when compromised by DC Sources*

in this security class that can be integrated on a chip with today's available technology, and it can have a wide range of related applications [59-72].

*1.1. The KLJN key exchange system*

The Kirchhoff-Law-Johnson-Noise (KLJN) secure key distribution scheme [41-57] is composed of two communicators (Alice and Bob) connected by a wire, see Fig. 1. Each communicator is constructed from a switch that at the beginning of bit exchange period randomly connects one of the two resistors, $R_L$ or $R_H$ ($R_L < R_H$), to the wire. Hence, the KLJN system has four different connected resistor situations, LL, HH, HL, and LH, where the first letter refers to Alice's connected resistor and the second letter to Bob's.

During the bit exchange period (BEP), Alice and Bob measure the current and voltage on the wire, and use their resistance value, Kirchhoff's law and the Johnson noise formula to determine the value of the connected resistance at the other end. On the other hand, an eavesdropper, whom we shall call Eve, can deduce only the resistance values of the connected resistors but not their location, unless the resistances are equal. Hence, the LH and HL cases are *secure situations*. Alice and Bob, who know also the locations of these resistance values, publicly agree about the bit value interpretation of the LH and HL situations. For example, they can agree that the HL and LH situations mean key bit values of 0 and 1, respectively.

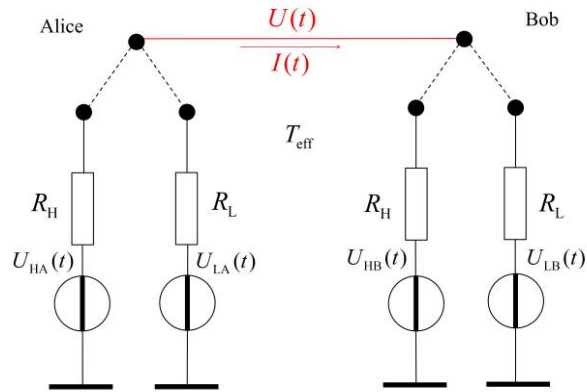

Fig 1. The core of the KLJN system. The two communicators at both Alice's and Bob's ends are connected by a wire. In the simplest system, Alice and Bob have identical pairs of resistors $R_L$ and $R_H$. They randomly choose and connect the selected resistors to the wire. The noise generators are denoted as $U_{HAn}(t)$, $U_{LAn}(t)$, $U_{HBn}(t)$ and $U_{LBn}(t)$, and they could be thermal, or artificial (enhanced for higher temperature).

Since its introduction [42] (see also feature [43]) and security proofs (e.g. [44]), many enhanced versions of the KLJN scheme have been developed for improved security, increased speed or reduced error probability, e.g. [41,44-49].

Practical applications have potential to introduce unconditional security, among others, for networks and smart grids [59-66]; computers, instruments, credit cards, key cards, PUF [67-69]; autonomic vehicle systems [70-71]; and accelerated IT-secure key exchange [72].

*1.2 Attacks*

The privacy properties of the KLJN system have intensively been scrutinized [41,42,73-85]; however the unconditional (information-theoretic) security of the system has not been successfully challenged. The attacks against the protocol can be categorized into three classes [1]:

(i) *Active attacks* where Eve either injects energy into the system or modifies the system to extract information. The literature discusses two types of this class of attacks:

- The man-in-the-middle (MITM) attack [41,73]. Eve cuts the wire and inserts two KLJN circuitries to imitate to Alice that she is Bob and to imitate to Bob that she is Alice. She can do that with resistors (Fig. 2), or twin noise current or voltage generators (Figs. 3 and 4). Eve then extracts the resistance values by using the Johnson's noise formula and Kirchhoff's law [73]. The ideal KLJN system is naturally protected against all these attacks by communicating and comparing the instantaneous voltages and currents in the wire via an authenticated public channel. A more advanced MITM attack uses circulators [46] but even this attack fails due to the communication between Alice and Bob and comparing their extracted bit values with their resistance values.

- The current injection attack where Eve injects a current in the wire and uses the measurement of the current distribution between the two directions (Alice/Bob) to estimate the values of the actual resistance values [42,74,87].

(ii) *Passive attacks* exploit existing information leaks in the non-ideal KLJN schemes due to non-ideal features (for example parasitic elements, such as resistance, capacitance, spurious voltages and currents, or inaccuracies) in the practical system [43,75-85]. Two recent passive attacks exploit the assumed existence of parasitic voltage sources (such as a ground loop) at one or both sides of the KLJN system [75-77,86]. The studies show that, with proper preparation [75-77], Alice and Bob are fully protected against these attacks, too.

(iii) Flawed attack attempts that have fundamental errors in their model, mathematics, physics, and/or experimental demonstration [86-92]. Even some of these attacks have been useful because they motivated discussions leading to a deeper understanding of the security [87] of the KLJN protocol. They lead even to the clarification of some fundamental questions, such as waves or no waves in cables driven by thermal noise, and their relation to Boltzmann's energy equipartition theorem [46,87,89].

In this paper, we study a new situation that is combined of former attack features: MITM [73] and the current injection attacks [42,74] in the presence of parasitic voltage sources (such as ground loop) [75-77] in the wire channel. Is it possible that the two former compromising factors synergistically help Eve to crack the system? We will see that the opposite of this is true in the MITM attack: the combined factors make it more difficult for Eve to keep her operation covert.

## 2. Man-in-the-middle attacks at parasitic DC sources

In [73], three different versions of the MITM attack were proposed:

(i) By inserted KLJN circuitries identical to Alice's and Bob's ones, see Fig. 2. The defense against this attack was the communication of measured instantaneous current values by Alice and Bob via an authenticated public channel. The noise currents at Alice and Bob become independent during the attack thus the eavesdropping is discovered instantaneously with very high probability [73].

(ii) By inserting twin noise current generators, see Fig. 4;

(iii) By inserting twin noise voltage generators, see Fig. 5.

Below we prove that the KLJN system compromised by parasitic DC sources [75-77] is not only secure against these attacks but, in their original form, some of these attacks help Alice and Bob to discover Eve in much simpler ways than that was proposed earlier [73].

*2.1 MITM attack by inserted KLJN circuitries*

It is obvious from the analysis of the circuitry in Fig. 2 that, whenever the parasitic DC sources of $U_{\text{DCA}}$ and $U_{\text{DCB}}$ are sufficiently different, Alice and Bob can use changes in the DC current to discover Eve. Then the DC situation will also change during the attack, and if the difference is greater than the rms value of the AC component of the wire voltage, Alice and Bob can successfully use simple time average to distinguish between the attack and no-attack situations. The original loop is broken in two separate loops thus the resultant DC loop voltage will be different in both new loops leading to different DC currents. Thus Alice and Bob can discover the attack even without comparing their instantaneous currents. The DC current they measure will change due to the attack thus *they can discover Eve without even communicating with the other party*. Thus, Alice and Bob need only to do a simple time averaging to uncover the attack.

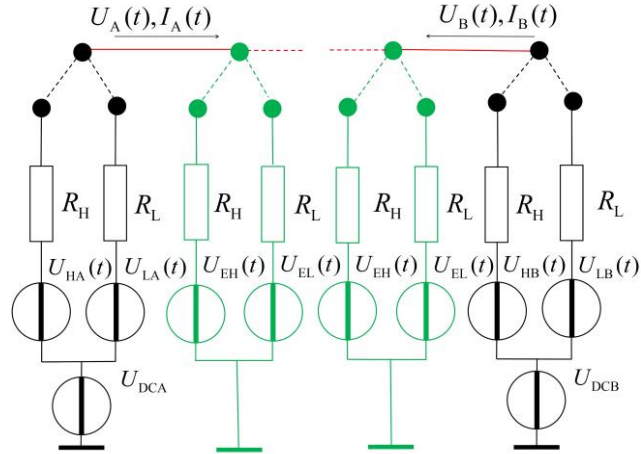

Fig. 2. Using the original [73] MITM attack. Eve cuts the wire and attaches two communicators to the KLJN system; one faces Alice's end and the other faces Bob's end. Thus two independent KLJN loops are created. Both communicators are identical with the communicators of Alice and Bob (except for the parasitic voltages). Even though Eve's noise voltage generators $U_{EHn}(t)$ and $U_{ELn}(t)$ have identical spectrum as the noise generators of Alice and Bob, their time functions are statistically independent. $I_A(t)$ and $U_A(t)$ are the instantaneous current and voltage at Alice's end, while $I_B(t)$ and $U_B(t)$ are the instantaneous current and voltage at Bob's end. By comparing the instantaneous values, Alice and Bob can discover the attack quickly [73]. However they can do that even without measuring the instantaneous currents: The DC current in their own lop will change due to the attack thus they can discover Eve without communicating with the other party and doing time averaging instead. For an improved attack, see Fig. 3.

Fig. 3 shows an *improved* MITM attack with inserted KLJN circuitries. By adding the proper DC voltage generators matching the situations at Alice's and Bob's ends, Eve's imitation of Alice and Bob is improved and Eve can stay hidden with the same probability as at the original MITM attack situation [73] because the DC current components at the two ends will remain the same. The defense against this attack is identical to the original protocol [73] thus, it requires the measurement and communication of measured instantaneous current values by Alice and Bob via an authenticated public channel. The noise (that is the AC) currents at Alice and Bob are statistically independent during the attack thus the eavesdropping is discovered instantaneously with very high probability. Mathematically speaking, the probability that Eve can stay hidden decays exponentially with time, and it is in the order of $10^{-20}$ over the bit exchange period at practical conditions [73].

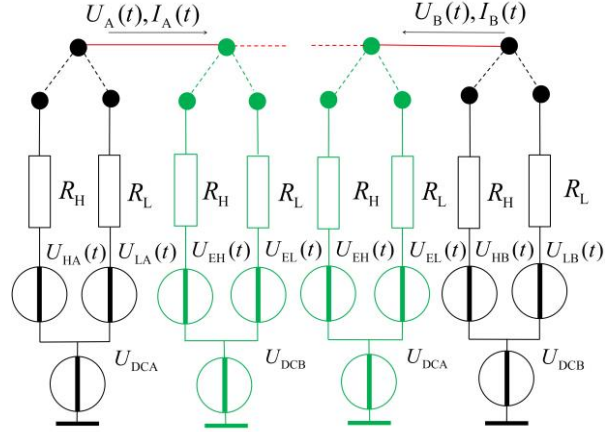

Fig. 3. Improved MITM attack with inserted KLJN circuitries. By adding the proper DC voltage generators Eve's imitation of Alice and Bob is improved and Eve can stay hidden with the same probability as in the original MITM attack [73].

*2.2. MITM attack by inserting twin noise current generators*

The man-in-the-middle attack with twin noise current generators, see Fig. 4, is the same as in the original MITM situation [73]. Even though Eve's injected currents are identical with $I_E(t)$ amplitude, due to Ohm's law, the instantaneous voltages $U_A(t)$ and $U_B(t)$ will be different most of the time because of the independence of the noise voltage generators of Alice and Bob, similarly to the original situation [73]. Thus Alice and Bob can quickly discover Eve by comparing their instantaneous voltage measurement data.

However, similarly to the scheme described in Section 2.1, whenever the parasitic DC sources of $U_{DCA}$ and $U_{DCB}$ are sufficiently different, the DC voltage situation in the wire will also change during the attack, and if the difference is greater than the rms value of the AC component of the wire voltage, Alice and Bob can successfully use simple time average over the bit exchange period to distinguish between the attack and no-attack situations. Here, Eve's solution (shown in Fig. 3) to fix this problem does not work because adding a voltage generator in serial to the current generators will obviously not change the voltage and current values in the loop. Thus the existence of the ground loop ultimately makes Eve's situation worse.

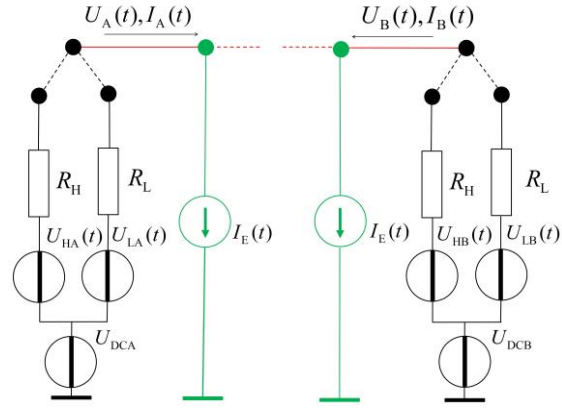

Fig. 4. The man-in-the-middle attack with twin noise current generators $I_E(t)$ of identical amplitudes [17]. Alice and Bub can compare the instantaneous voltage values via an authenticated public channel or, whenever the parasitic DC voltages are sufficiently different, utilize simple time averaging to uncover the attack.

## 2.3. MITM attack by inserting twin noise voltage generators

The man-in-the-middle attack with twin noise voltage generators, see Fig. 4, is the same as in the original MITM situation [73]. Even though Eve's injected currents are identical with $U_E(t)$ amplitude, due to Ohm's law the instantaneous currents $I_A(t)$ and $I_B(t)$ will be different at most of the time due to the independence of the noise voltage generators of Alice and Bob, similarly to the original situation [73]. Thus Alice and Bob can quickly discover Eve by comparing their instantaneous voltage measurement data [73].

However, similarly to the scheme described in Section 2.1, whenever the parasitic DC sources of $U_{DCA}$ and $U_{DCB}$ are sufficiently different, the DC current situation in the wire will also change during the attack and, if the difference is greater than the rms value of the AC component of the wire current, Alice and Bob can successfully use simple time average of the current over the bit exchange period to distinguish between the attack and no-attack situations.

*Active Attack against the KLJN scheme when compromised by DC Sources*

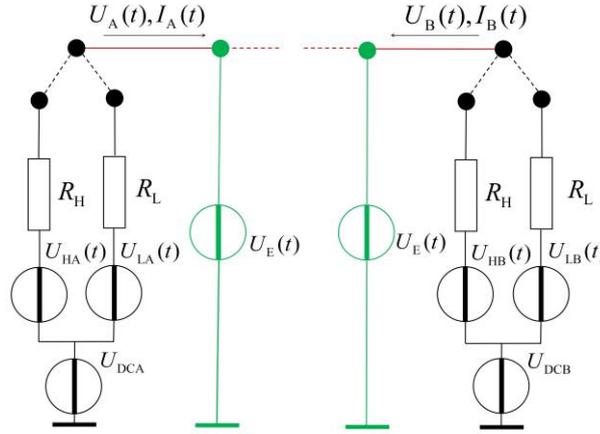

Fig. 5. The man-in-the-middle attack with twin noise voltage generators $U_E(t)$ of identical amplitudes. For other definitions, see the captions of Fig. 1 and Fig. 2. Alice and Bob can discover the attack even without comparing their instantaneous currents. The DC current in their lop will change due to the attack thus they can discover Eve without communicating with the other party. For an improved attack, see Fig. 6.

Fig. 6 shows a somewhat improved attack where Eve is attempting to compensate the DC voltage differences at the two ends by forcing a proper DC voltage component on the wire. However she can do this deterministically only during emulating the HH or LL situation when the DC voltage on the wire is the average of the parasitic DC voltage components of the two ends. At these situations the bits are discarded thus Eve des not gain any advantage.

In the secure (LH and HL) cases, Eve has only 0.5 success probability to eliminate the DC problem. For example, toward Bob, she can fix the DC voltage to imitate the DC voltage drop at the LH situation and wait until Bob randomly choses the H resistor. The problem with this approach is that, if Bob choses L instead, he will detect that the DC voltage is the wrong value. Thus this attack improvement is strongly limited as works only in half of the cases thus Alice and Bob can discover the attack, on the average, in two secure bit exchange periods even without public communications.

On the other hand, the original defense method [73], that is, the public comparison of the instantaneous voltage measurement data of Alice and Bob always uncovers Eve with very high probability in a very short time. Her probability to be able to stay hidden decays exponentially and reaches miniscule values [73] as mentioned above in Section 2.1.

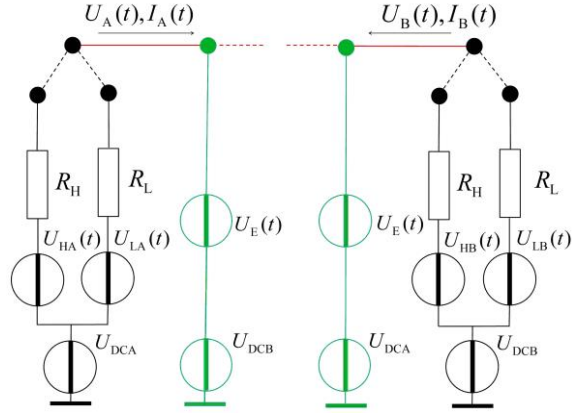

Fig. 6. Improved twin voltage generator attack. By adding the proper DC voltage generators Eve's imitation of Alice and Bob is improved and Eve can stay hidden with the same probability as in the original situation.

### 3. Current Injection Attack at parasitic DC sources

During the current injection attack [42,74,87], see Fig. 7, Eve injects a small noise current in the line and measures the cross-correlation of this current with the current in the wire. Due to Kirchhoff's node law, the cross-correlation is greater at that side of the attack point where the resistance is lower because the injected current is distributed according the conductance of the terminations of the wire. The efficient defense against this attack is the comparison of instantaneous current amplitudes by Alice and Bob which were formerly proposed [42], analyzed [7,87] and tested [50]. The goal of our current study is to clarify:

(a) If Eve has any advantage or disadvantage from the parasitic DC sources and the related loop current;

(b) If the former defense protocol [42,74,87] still works.

The answer to the above questions is straightforward:

(i) The parasitic DC current is a DC current component. Thus it is statistically independent from the AC current components (including Eve's ones) in the wire. Thus the parasitic Dc currents have zero contribution to Eve's cross-correlation measurement. Moreover, by using a current generator, Eve can conduct the same attack, without any advantage or disadvantage, as earlier.

(ii) If Alice and Bob compare the instantaneous current amplitudes of the AC components of their measurement, the detection situation (and resolution) also remains the same as earlier.

*Active Attack against the KLJN scheme when compromised by DC Sources*

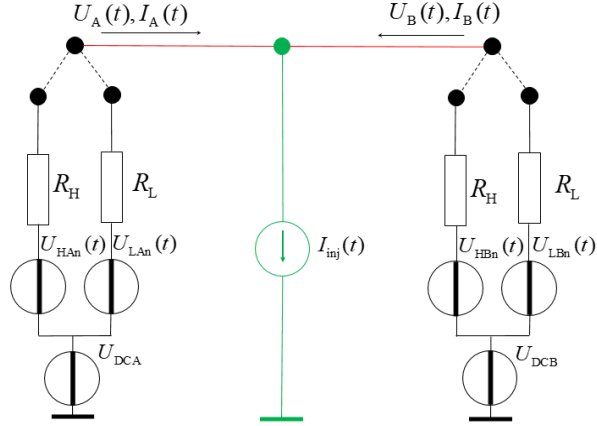

Fig. 7. The current injection attack against the KLJN scheme, where $I_{inj}(t)$ is the injected current. For other definitions, see the captions of Fig.1 and Fig. 2. The existence of DC sources does not influence the distribution of injected AC currents thus the security level remains the same as without them. Alice and Bob discover the attack by comparing their instantaneous AC current amplitude.

In conclusion, the parasitic DC sources do not influence the current injection attack and its defense.

## 4. Conclusion

We proved a practical KLJN system that is possibly compromised by parasitic serial DC sources remains secure against the man-in-the-middle attacks and the current injection attack. In each case, Eve's probability of success to stay hidden is less or equal to the idealistic situation with no parasitic sources [73]. Thus launching these attacks against the KLJN scheme remains non-feasible.